\documentclass[conference,a4paper]{IEEEtran}
\IEEEoverridecommandlockouts

\usepackage{graphicx}
\usepackage{caption}
\usepackage{subcaption}
\usepackage{amsmath,amssymb,amsthm,amsfonts}
\usepackage{algorithm}
\usepackage{algpseudocode}
\usepackage{mathtools}
\usepackage{hyperref}
\usepackage{stmaryrd}

\usepackage{caption}
\usepackage{subcaption}
\usepackage[table]{xcolor}
\usepackage{xfrac}

\newtheorem{theorem}{Theorem}[section]

\usepackage[framemethod=tikz]{mdframed}

\makeatletter
\def\ProblemSpecBox{
  \@ifnextchar[\ProblemSpecBox@opt{\ProblemSpecBox@noopt}}
\def\ProblemSpecBox@opt[#1]#2{
  \protected@edef\@currentlabelname{#1}
  \protected@edef\@currentlabel{#1}
  \begin{mdframed}[
    innerleftmargin=5pt,
    innerrightmargin=5pt,
    innertopmargin = 5pt,
    innerbottommargin=5pt,
    skipabove=\dimexpr\topsep+\ht\strutbox\relax,
    roundcorner=2pt,
    frametitle={#2},
    frametitlerule=true,
    backgroundcolor=gray!7,
    frametitlebackgroundcolor=gray!20]
}
\def\ProblemSpecBox@noopt#1{
  \ProblemSpecBox@opt[#1]{#1}
}
\def\endProblemSpecBox{
  \end{mdframed}
}
\makeatother

\usepackage{array}
\usepackage{booktabs}
\usepackage{graphicx}
\usepackage{threeparttable}
\usepackage{wasysym} 
\usepackage{multirow}

\begin{document}
%
\title{Poster: \textsc{\mdseries RandGener}: Distributed Randomness Beacon from Verifiable Delay Function}

\author{\IEEEauthorblockN{Arup Mondal}
\IEEEauthorblockA{Ashoka University \\
Sonipat, Haryana, India \\
arup.mondal\_phd19@ashoka.edu.in}
\and
\IEEEauthorblockN{Ruthu Hulikal Rooparaghunath}
\IEEEauthorblockA{Vrije Universiteit \\
Amsterdam, The Netherlands
 \\
r.rooparaghunath@student.vu.nl}
\and
\IEEEauthorblockN{Debayan Gupta}
\IEEEauthorblockA{Ashoka University \\
Sonipat, Haryana, India \\
debayan.gupta@ashoka.edu.in}}
\maketitle

\begin{abstract}
Buoyed by the excitement around secure decentralized applications, the last few decades have seen numerous constructions of distributed randomness beacons (DRB) along with use cases; however, a secure DRB (in many variations) remains an open problem. 
We further note that it is natural to want some kind of reward for participants who spend time and energy evaluating the randomness beacon value -- this is already common in distributed protocols.

In this work, we present \textsc{RandGener}, a \textit{novel} $n$-party \textit{commit-reveal-recover} (or \textit{collaborative}) DRB protocol with a novel \textit{reward and penalty mechanism} along with a set of realistic guarantees. 
We design our protocol using trapdoor watermarkable verifiable delay functions in the RSA group setting (without requiring a trusted dealer or distributed key generation).

\end{abstract}



\section{Introduction}

A \textit{randomness beacon}~\cite{DBLP:journals/sigact/000183} is an ideal functionality that continuously publishes independent random values which no party can predict or manipulate; critically, this value must be efficiently verifiable by anyone. 
A \textit{Distributed Randomness Beacon} (DRB) protocol allows a set of participants to jointly compute a continuous stream of randomness beacon outputs. A secure DRB protocol should satisfy the following properties, outlined in~\cite{DBLP:conf/ndss/SchindlerJHSW21,DBLP:journals/iacr/ChoiATB23,DBLP:journals/iacr/LenstraW15}:
\begin{itemize}
    \item [\textcolor{blue}{(1)}] \textit{Liveness/availability}: participants should not be able to prevent the progress of random beacon computation, 
    \item[\textcolor{blue}{(2)}] \textit{Guaranteed output delivery}: adversaries should not be able to prevent honest  participants in the protocol from obtaining a random beacon output,
    \item[\textcolor{blue}{(3)}] \textit{Bias-resistance}: no participants should be able to influence future random beacon values to their advantage,
    \item[\textcolor{blue}{(4)}] \textit{Public verifiability}: as soon as a random beacon value is generated, it can be verified by anyone independently using only public information, and
    \item[\textcolor{blue}{(5)}] \textit{Unpredictability}: participants should not be able to predict the future random beacon values. 
\end{itemize}

We introduce two new desirable properties for DRB protocols: \textcolor{blue}{(6)} a \textit{\textbf{reward mechanism}}, which incentivizes participants who invest time and energy in evaluating the randomness beacon value by rewarding their effort, and \textcolor{blue}{(7)} a \textbf{\textit{penalty mechanism}}, which discourages inadequate participation, incorrect information or cheating by applying penalties for participants who engage in those actions.

Our $n$-party distributed randomness beacon protocol, \textsc{RandGener} demonstrates a method of claiming ``ownership'' of a randomness beacon value evaluation in each round of the protocol's execution. This is done by attaching a ``watermark'' of computing participants to the result of the evaluation in order to reward  corresponding participants for their contribution.

\textit{Our contributions are summarised as follows:}

\begin{itemize}
    \item We extend watermarkable VDF (wVDF) defined in~\cite{DBLP:conf/eurocrypt/Wesolowski19} by formally defining a new type called \textit{\textbf{trapdoor}} wVDF.  Furthermore, we demonstrate a construction using Wesolowski \cite{DBLP:conf/eurocrypt/Wesolowski19} and Pietrzak's \cite{DBLP:journals/iacr/Pietrzak18a} scheme.
    
    \item We construct \textsc{RandGener}, an efficient $n$-party \textit{commit-reveal-recover} (or \textit{collaborative}) distributed randomness beacon protocol with a novel \textit{\textbf{reward mechanism}} and \textit{\textbf{penalty mechanism}} using a trapdoor wVDF. Our protocol does not require any trusted (or expensive) setup and proves that it provides the desired security properties. 
\end{itemize}

\subsubsection*{Brief Relevant Work}

A \textit{commit-reveal} is a classic approach proposed in~\cite{DBLP:journals/sigact/000183}. First, all participants publish a commitment $y_i = \textsf{Commit}(x_i)$  to a random value $x_i$.  Next, participants reveal their $x_i$ values, resulting in $R = \textsf{Combine}(x_1, \dots, x_n)$ for some suitable combination function (such as an exclusive-or or a cryptographic hash).

However, the output can be biased by the last participant to open their commitment (referred to as a \textit{last-revealer attack}), since the last participant, by knowing all other commitments $x_i$, can compute $R$ early.

A very different approach to constructing DRBs uses time-sensitive cryptography (TSC), specifically using delay functions to prevent manipulation. The simplest example is Unicorn~\cite{DBLP:journals/iacr/LenstraW15}, a one-round protocol in which participants directly publish (within a fixed time window) a random input $x_i$. The result is computed as $R = \textsf{TSC}(\textsf{Combine}(x_1, \dots, x_n))$. However, the downside of the Unicorn~\cite{DBLP:journals/iacr/LenstraW15} is that a delay function must be computed for every run of the protocol.
Recently, Choi et al.~\cite{DBLP:journals/iacr/ChoiATB23} introduced the Bicorn family of DRB protocols, which retain the advantages of Unicorn~\cite{DBLP:journals/iacr/LenstraW15} while enabling efficient computation of the result (with no delay) if all participants act honestly. Yet, as stated in~\cite{DBLP:journals/iacr/ChoiATB23}, all Bicorn variants come with a fundamental security caveat, i.e., the \textit{last revealer prediction attack}: if participant $P_i$ withholds their $x_i$ value, but all others publish, then participant $P_i$ will be able to simulate efficiently and learn $R$ quickly (\textit{optimistic case}), while  honest participants will need to execute the force open and compute the delay function to complete before learning $R$ (\textit{pessimistic case}). Similarly, a coalition of malicious participants can share their $x$ values and privately compute $R$. 
Nevertheless, \textit{none} of the existing delay-cryptography-based commit-reveal-recover style DRB protocols provide a reward/penalty mechanism to regulate the behaviour of corrupted participants. In this work, we propose an efficient $n$-party commit-reveal-recover (or collaborative) DRB protocol with a novel reward and penalty mechanism based on the trapdoor wVDF.

\begin{table}[!ht]
    \centering
    \scriptsize
	\caption{Comparison collaborative DRB schemes.}
	\label{table:drbschemes}
	\newcommand*\rot[1]{\hbox to1em{\hss\rotatebox[origin=br]{-60}{#1}}}
	\newcommand*\feature[1]{\ifcase#1 $\square$\or$\boxtimes$\or$\blacksquare$\or\Circle\or\LEFTcircle\or\CIRCLE\or$\rhd$\or$\lhd$\or$\checkmark$\or$\times$\or \fi}
	\newcommand*\e[9]{\feature#1&\feature#2&\feature#3&\feature#4&\feature#5&\feature#6&\feature#7&\feature#8&\feature#9}
	\newcommand*\f[3]{\feature#1&\feature#2&\feature#3}
	\makeatletter
	\newcommand*\ex[7]{#1\tnote{#2}&#3&#4&#5&#6&\f#7&\expandafter\e\@firstofone
	}
	\makeatother
	\newcolumntype{E}{c@{}c@{}c@{}c@{}c@{}c@{}c@{}c@{}c}
	\newcolumntype{F}{c@{}c@{}c}
	\begin{threeparttable}
		\begin{tabular}{@{}l|c|cc|c F E@{}}
			\toprule
            \rowcolor{gray!30}
			Paper & \shortstack{Crypto \\ Primit.} & \shortstack{Comp. \\ Cost} & \shortstack{Comm. \\ Cost} & \shortstack{Fault \\ Toler.} & \multicolumn{3}{c}{\shortstack{Crypto \\ Model}} & \multicolumn{9}{c}{Features}\\
			\midrule

			&& \rot{}
			& \rot{}
			& \rot{}
			& \rot{\shortstack{Network Model}}
			& \rot{\shortstack{Setup Assumption}}
			& \rot{\shortstack{Trusted Setup Req.}}
			%
			& \rot{\shortstack{Adaptive Adversary}}
			& \rot{\shortstack{Liveness/Availability}}
			& \rot{\shortstack{Bias Resistance}}
			& \rot{\shortstack{Fairness}}
			& \rot{\shortstack{GOD}}
			& \rot{\shortstack{Scaleability}}
			& \rot{\shortstack{Unpredictability}}
			& \rot{\textcolor{black}{\textbf{Reward}}}
			& \rot{\textcolor{black}{\textbf{Penalty}}} \\
			\midrule
			
			\ex{\cite{DBLP:journals/iacr/LenstraW15}}{} {Sloth} {$O(n)$} {$O(1)$} {$\frac{(n-1)}{n}$} {248} {988889899} \\

			\ex{\cite{DBLP:conf/ndss/SchindlerJHSW21}}{} {tVDF} {$O(n^2)$} {$O(1)$} {$\frac{n}{2}$} {049} {888888999} \\
			
			\ex{\cite{DBLP:journals/iacr/ChoiATB23}}{} {VDF} {$O(n^2)$} {$O(1)$} {$n-1$} {149} {888888899} \\
			
			\midrule\midrule
			
			\ex{Ours}{} {twVDF} {$O(n^2)$} {$O(1)$} {$n-1$} {149} {888888888} \\
			
			\bottomrule
		\end{tabular}
		
		\begin{tablenotes}
			\item \feature0 denotes the ``asynchronous'' network model; \feature1 denotes the ``partial synchronous'' network model; \feature2 denotes the ``synchronous'' network model.
			\feature4 denotes the ``Common Reference String'' setup assumption; 
			\feature8 denotes provide the property; \feature9 denotes does not provide the property. GOD -- Guaranteed Output Delivery.
   
		\end{tablenotes}
	\end{threeparttable}
\end{table}

\section{Technical Preliminaries}

\subsubsection*{Basic Notation}
Given a set $\mathcal{X}$, we denote $x \overset{\$}{\leftarrow} \mathcal{X}$ as the process of sampling a value $x$ from the uniform distribution on $\mathcal{X}$. Supp($\mathcal{X}$) denotes the support of the distribution $\mathcal{X}$. 

We denote the security parameter by $\lambda \in \mathbb{N}$. A function \texttt{negl}: $\mathbb{N} \rightarrow \mathbb{R}$ is negligible if it is asymptotically smaller than any inverse-polynomial function. Namely, for every constant $\epsilon > 0$ there exists an integer $N_{\epsilon}$ for all $\lambda > N_{\epsilon}$ such that $\texttt{negl}(\lambda) \leq \lambda^{-\epsilon}$. 

\subsubsection*{Number Theory}
We assume that $N=p \cdot q$ is the product of two large secret and \textit{safe} primes and $p \neq q$. We say that $N$ is a strong composite integer if $p = 2p' + 1$ and $q = 2q' +1$ are safe primes, where $p'$ and $q'$ are also prime. We say that $\mathbb{Z}_N$ consists of all integers in $[N]$ that are relatively prime to $N$ (i.e., $\mathbb{Z}_N = \{x \in \mathbb{Z}_N: \texttt{gcd}(x, N)=1\}$).

\subsubsection*{Repeated Squaring Assumption}
The repeated squaring assumption~\cite{DBLP:journals/iacr/Pietrzak18a} roughly states that there is no parallel algorithm that can perform $T$ squarings modulo an integer $N$ significantly faster than just doing so sequentially, assuming that $N$ cannot be factored efficiently, or in other words \texttt{RSW} assumption implies that factoring is hard. More formally, no adversary can factor an integer $N = p \cdot q$ where $p$ and $q$ are large secret and ``safe'' primes~\cite{DBLP:journals/iacr/Pietrzak18a}.  A repeated squaring \texttt{RSW = (Setup, Sample, Eval)} is defined below. Moreover, we define a trapdoor evaluation \texttt{RSW.tdEval} (to enable \textit{fast} repeated squaring evaluation), from which we can derive an actual output using trapdoor in $\textsf{poly}(\lambda)$ time.

\begin{mdframed}
\scriptsize
    $\bullet$ $N \leftarrow \texttt{RSW.Setup}(\lambda)$ : Output $\textsf{pp} = (N)$ where $N = p \cdot q$ as the product of two large ($\lambda$-bit) randomly chosen secret and safe primes $p$ and $q$.
    
    \noindent
    $\bullet$ $x \leftarrow \texttt{RSW.Sample}(\textsf{pp})$ : Sample a random instance $x$.
    
    \noindent
    $\bullet$ $y \leftarrow \texttt{RSW.Eval}(\textsf{pp}, T, x)$ : Output $y = x^{2^{T}} \mod N$ by computing the $T$ sequential repeated squaring from $x$. 
    
    \noindent
    $\bullet$ $y \leftarrow \texttt{RSW.tdEval}(\textsf{pp}, \textsf{sp} = \phi(N), x)$ : To compute $y = x^{2^{T}} \mod N$ \textit{efficiently} using the trapdoor as follows:
    \textcolor{blue}{\textit{\textbf{(i)}}} Compute $v = 2^T \mod \phi(N)$. \textbf{Note:} $(2^T \mod \phi(N)) \ll 2^T$ for large $T$. \textcolor{blue}{\textit{\textbf{(ii)}}} Compute $y = x^v \mod N$. \textbf{Note:} $x^{2^{T}} \equiv x^{(2^T \mod \phi(N))} \equiv x^v \pmod{N}$.
\end{mdframed}

\section{Verifiable Delay Function}

A \textit{verifiable delay function} (VDF), introduced by Boneh et al.~\cite{DBLP:journals/iacr/BonehBBF18}, is a special type of delay function $f$ characterized by a time-bound parameter $T$ and the following three properties: 
\textcolor{blue}{\textbf{(\textit{i})}} \textit{$T$-sequential function}: The function $f$ can be evaluated in sequential time $T$, but it should not be possible to evaluate $f$ significantly faster than $T$ even with parallel processing.
\textcolor{blue}{\textbf{(\textit{ii})}} \textit{Unique output}: The function $f$ produces a unique output, which is efficiently and publicly \textcolor{blue}{\textbf{(\textit{iii})}} \textit{Verifiable} (in time that is essentially independent of $T$) - meaning that the function $f$ should produce a proof $\pi$ which convinces a verifier that the function output has been correctly computed.

Wesolowski~\cite{DBLP:conf/eurocrypt/Wesolowski19} first describes a trapdoor VDF (tVDF) as a modified and extended version of traditional VDFs~\cite{DBLP:journals/iacr/BonehBBF18} such that the \texttt{Setup} algorithm, in addition to the public parameters \textsf{pp}, outputs a trapdoor or secret parameter \textsf{sp} to the party invoking the \texttt{Setup} algorithm. This parameter \textsf{sp} is kept secret by the invoker, whereas \textsf{pp} is published. Furthermore, using the trapdoor evaluation \texttt{tdEval} and trapdoor proof generation \texttt{tdProve} (by enabling \textit{fast} computations), the secret parameter-holding participants can derive an actual output and the proof of correctness in $\textsf{poly}(\lambda)$ time. Parties without knowledge of the trapdoor, as in the traditional VDF case, can still compute the output and proof of correctness by executing \texttt{Eval} and \texttt{Prove}. However, it requires $T$-sequential steps to do so.
For the purpose of our distributed random beacon protocol, we require and define a trapdoor watermarkable VDF (twVDF). In this case, we use the same trapdoor evaluation \texttt{tdEval}, but we generate a watermarked proof of correctness using \texttt{tdProve} by embedding a watermark of the evaluator.

In Algorithm~\ref{algo:vdf}, we provide details for the formal construction of watermarkable verifiable delay function \texttt{VDF} using Wesolowski~\cite{DBLP:conf/eurocrypt/Wesolowski19} and Pietrzak's~\cite{DBLP:journals/iacr/Pietrzak18a} scheme, consisting of algorithms (\texttt{Setup, Sample, Eval, Prove, Verify}) with a trapdoor watermarkable VDF evaluation \texttt{tdEval} and proof generation \texttt{tdProve}.

\begin{ProblemSpecBox}[1]{\scriptsize Algorithm 1: Trapdoor Watermarkable VDF using Wesolowski and Pietrzak}
\label{algo:vdf}
\scriptsize


    \smallskip
    $\bullet$ \verb|VDF.Setup|($\lambda$)
    \begin{enumerate}
        \item Call and generate $N \leftarrow \texttt{RSW.Setup}(\lambda)$
        \item A cryptographically secure $\lambda$-bit hash function $\mathcal{H}_{\textsf{prime}}$ or $\mathcal{H}_{\textsf{random}}$. 
        \item Generate a time-bound parameter $T$.
        \item \textbf{return} $\textsf{pp} = (N, T, \mathcal{H})$.
    \end{enumerate}

	$\bullet$ \verb|VDF.Sample|($\textsf{pp}$)
	\begin{enumerate}
		\item Generate an input $x \in \mathbb{Z}_N^* \leftarrow \texttt{RSW.Sample}(\textsf{pp})$
	\end{enumerate}
    
    $\bullet$ \verb|VDF.Eval|($\textsf{pp},x$) 
    \begin{enumerate}
         \item Compute $y = x^{2^T} \mod N \in \mathbb{Z}_N^*$ using $\texttt{RSW.Eval}(\textsf{pp}, T, x)$
        \item Generate an advice string $\alpha$.
        \item \textbf{return} ($y, \alpha$).
    \end{enumerate}

    $\bullet$ \verb|VDF.tdEval|($\textsf{pp},x$) 
    \begin{enumerate}
    	\item Compute group order $\phi(N) = (p-1) \cdot (q-1)$ using trapdoor $(p, q)$.
    	\item Compute $y = x^{2^T} \mod N$ using $\texttt{RSW.tdEval}(\textsf{pp}, \textsf{sp} = \phi(N), x)$.
	
    	\item Generate an advice string $\alpha$.
    	\item \textbf{return} ($y, \alpha$).
    \end{enumerate}
	
	\smallskip
	\noindent\dotfill \textcolor{black}{\textsf{~\textbf{Using Wesolowski's~\cite{DBLP:conf/eurocrypt/Wesolowski19} Scheme}~}} \dotfill
	\smallskip
 
    $\bullet$ \verb|VDF.Prove|($\textsf{pp},x,\mu,y,\alpha,T$)
    \begin{enumerate}
    	\item Generate a prime $l = \mathcal{H}_{\textsf{prime}}(x \parallel y \parallel \mu)$ \footnote{\scriptsize Sampled uniformly from $\textsf{Prime}(\lambda)$}
    	\item Compute the proof $\pi_{\mu} = x^{\lfloor{2^{T}/l}\rfloor} \pmod{N}$ \footnote{$\mu$ is an evaluator's watermark}
    	and \textbf{return} $\pi_{\mu}$.
    \end{enumerate}

	$\bullet$ \verb|VDF.tdProve|($\textsf{pp},x,\mu,y,\alpha,T$)
	\begin{enumerate}
		\item Generate a prime $l = \mathcal{H}_{\textsf{prime}}(x \parallel y \parallel \mu)$
		\item Compute group order $\phi(N) = (p-1) \cdot (q-1)$ using trapdoor $(p, q)$.
		\item Compute proof $\pi_{\mu} = x^{(\lfloor{2^{T}/l}\rfloor \mod \phi(N))} \pmod{N}$
		and \textbf{return} $\pi_{\mu}$.
	\end{enumerate}
    
    $\bullet$ \verb|VDF.Verify|($\textsf{pp},x,\mu,y,\pi_{\mu},T$)
    \begin{enumerate}
        \item Generate a prime $l = \mathcal{H}_{\textsf{prime}}(x \parallel y \parallel \mu)$
        \item $r = 2^{T} \mod l$
        \item \textbf{return} \textsf{accept} \textbf{if} $(\pi_{\mu}^l \cdot x^r) \mod N = y$, otherwise \textsf{reject}
    \end{enumerate}
	
	\smallskip
	\noindent\dotfill \textcolor{black}{\textsf{~\textbf{Using Pietrzak's~\cite{DBLP:journals/iacr/Pietrzak18a} Scheme}~}} \dotfill
	\smallskip
 
	$\bullet$ \verb|VDF.Prove|($\textsf{pp},x,\mu,y,\alpha,T$)
	\begin{enumerate}
		\item Compute $u = x^{2^{T/2}} \mod N$
		\item Generate a random $r = \mathcal{H}_{\textsf{random}}(x \parallel T/2 \parallel y \parallel u \parallel \mu)$ \footnote{\scriptsize Sampled uniformly from $\{1, 2, \dots, 2^{\lambda}\}$}
		\item Compute $x = x^{r} \cdot u \mod N$ and $y = u^{r} \cdot y \mod N$
		\item Proof $\pi_{\mu} = u \; \cup \; \texttt{VDF.Prove}(\textsf{pp}, x, \mu, y, \alpha, T/2)$
		and \textbf{return} $\pi_{\mu}$.
	\end{enumerate}

	$\bullet$ \verb|VDF.tdProve|($\textsf{pp},x,\mu,y,\alpha,T$)
	\begin{enumerate}
		\item Compute group order $\phi(N) = (p-1) \cdot (q-1)$ using trapdoor $(p, q)$.
		\item Compute $u = x^{(2^{T/2} \mod \phi(N))} \mod N$
		\item Generate a random $r = \mathcal{H}_{\textsf{random}}(x \parallel T/2 \parallel y \parallel u \parallel \mu)$
		\item Compute $x = x^{(r \mod \phi(N))} \cdot u \mod N$ and $y = u^{(r \mod \phi(N))} \cdot y \mod N$
		\item Proof $\pi_{\mu} = u \; \cup \; \texttt{VDF.tdProve}(\textsf{pp}, x, \mu, y, \alpha, T/2)$
		and \textbf{return} $\pi_{\mu}$.
	\end{enumerate}

	$\bullet$ \verb|VDF.Verify|($\textsf{pp},x,\mu,y,\pi_{\mu}, T$)
	\begin{enumerate}
		\item Generate a random $r = \mathcal{H}_{\textsf{random}}(x \parallel T/2 \parallel y \parallel u \parallel \mu)$
		\item Compute $x = x^{r} \cdot u \mod N$ and $y = u^{r} \cdot y \mod N$
		\item Call \texttt{VDF.Verify}($\textsf{pp},x,\mu,y,\pi_{\mu}, T/2$)
		\item \textbf{return} \textsf{accept} if $T = 1$ check $y = x^2 \mod N$, otherwise \textsf{reject}.
	\end{enumerate}
\end{ProblemSpecBox}

\section{\textsc{RandGener} Protocol Design}

In this section, we present our \textsc{RandGener} protocol, a $n$-party distributed randomness beacon protocol \texttt{DRB = (Setup, VerifySetup, Gen)}. The construction details are in Algorithm~\ref{algo:ranbec} using our trapdoor watermarkable VDF.

\begin{ProblemSpecBox}[2]{\scriptsize Algorithm 2: \textsc{\mdseries RandGener}: Distributed Randomness Beacon Protocol}
\label{algo:ranbec}
\scriptsize
\textbf{Input:} A globally agreed security parameter $\lambda$, a set of participants $\mathcal{P} = \{P_1, P_2, \dots, P_n\}$, a set of public parameters $\mathcal{PP} = \{\textsf{pp}_1, \textsf{pp}_2, \dots, \textsf{pp}_n\}$, a time-bound parameter $T$, an initial random beacon value $R_0$ (it becomes available to all parties running the protocol after the setup is completed at approximately the same time), and two cryptographically secure $\lambda$-bit hash functions: (i) $\mathcal{H}_{\textsf{randToinput}}$ -- mapping a random value to the input space of the VDF, and (ii) $\mathcal{H}_{\textsf{inputTorand}}$ -- mapping a VDF output to a random value.\\
\textbf{Output}: The randomness beacon value $R_1, R_2, \dots, R_{\infty}$ for that round of the protocol.
        
        \medskip
		$\bullet$ \texttt{DRB.Setup}$(\lambda)$
		\begin{enumerate}
			\item $\forall i$ $P_i \in \mathcal{P}$ locally generate a public parameter $\textsf{pp}_i = (N_i, T, \mathcal{H}) = \texttt{VDF.Setup}(\lambda)$.
			
			\item $\forall i$ $P_i \in \mathcal{P}$ run the zero-knowledge protocol for proving that a
			known $N_i$ is the product of two safe primes and the protocol ``proving the knowledge of a discrete logarithm that lies in a given range'' to show that the prime factors $p_i$ and $q_i$ are $\lambda$-bits each. Let $\pi_{N_i}$ denote the resulting proof obtained by running both protocols non-interactively using the Fiat-Shamir heuristic.
			
			\item \textbf{Broadcast}  ($\mathcal{PP} = \{\textsf{pp}_1, \dots, \textsf{pp}_n\}$, $\Pi = \{\pi_{N_1}, \dots, \pi_{N_n}\}$). 
		\end{enumerate}
		
		$\bullet$ \verb|DRB.VerifySetup|($\mathcal{PP}, \Pi$)
		\begin{enumerate}
			\item For each public parameter $\textsf{pp}_i \in \mathcal{PP}$ and a corresponding proof $\pi_{N_i} \in \Pi$, \textbf{return} \textsf{accept} if the validity of $\textsf{pp}_i$ can be successfully checked by using the verification procedures corresponding to the proof techniques used in \texttt{DRB.Setup} algorithm as specified in~\cite{DBLP:conf/ndss/SchindlerJHSW21}, otherwise \textbf{return} \textsf{reject}.
		\end{enumerate}
	    
		$\bullet$ \texttt{DRB.Gen}$(\mathcal{PP}, T, R_0)$
		\begin{enumerate}
			\item Set $r \leftarrow 1$. 
		    
		    \smallskip
		    \noindent\dotfill \textcolor{black}{\textsf{~\textbf{Commit}~}} \dotfill $~\textbf{deadline}~ T_0$
		    \smallskip
			\item Compute $x_r \leftarrow \mathcal{H}_{\textsf{randToinput}}(R_{r-1})$. 
			\item Generate a random input $x'_{r,i}$
            \item Compute and publish $x_{r,i} \leftarrow \mathcal{H}(x'_{r,i}||x'_r)$ \Comment{Broadcast}
			
			\smallskip
			\noindent\dotfill \textcolor{black}{\textsf{~\textbf{Reveal}~}} \dotfill $~\textbf{deadline}~ T_1$
			\smallskip
			\item For each participant $P_i \in \mathcal{P}$ in parallel
			\begin{enumerate}
                \item Compute $(y_{r,i}, \alpha_{r,i}) \leftarrow \texttt{VDF.tdEval}(\textsf{pp}_i, x_{r, i}, T)$.
                \item Compute $\pi_{r,i} \leftarrow \texttt{VDF.tdProve}(\textsf{pp}_i, x_{r,i}, \mu_i, y_{r,i}, \alpha_{r,i}, T)$.
                \item Publish $(y_{r,i}, \pi_{r,i})$. \Comment{Broadcast}
			\end{enumerate}
			\smallskip
			\noindent\dotfill \textcolor{black}{\textsf{~\textbf{Finalize}~}} \dotfill
			\smallskip
			
			\item For each participant $P_i \in \mathcal{P}$, verify $(x_r, y_{r,i}, \pi_{r,i})$
			\begin{enumerate}
			    \item If $\texttt{VDF.Verify}(\textsf{pp}_i, x_{r,i}, y_{r,i}, \pi_{r,i}, T) = \textsf{reject}$ or $x_{r,i}$ was not published by $T_1$, then remove participant $P_i$ and add $\tilde{\mathcal{P}} \leftarrow \tilde{\mathcal{P}} \cup P_i$.
			\end{enumerate}
			
			\item For \textbf{all} $P_i \in \mathcal{P}$, If $\texttt{VDF.Verify}(\textsf{pp}_i, x_{r,i}, y_{r,i}, \pi_{r,i}, T) = \textsf{accept}$ 
			\begin{enumerate}
			    \item Compute $y_r = \prod_{P_i \in \mathcal{P}} y_{r, i}$ \textbf{\textcolor{blue}{\Comment{Optimistic case}}}
			    \item Optionally, a proof $\pi_{y_r}$ can be compute to enable verification of $y_r$.
			\end{enumerate}
			
			\smallskip
			\noindent\dotfill \textcolor{black}{\textsf{~\textbf{Recover}~}} \dotfill
			\smallskip
			\item For each participant $P_j \in \tilde{\mathcal{P}}$ in parallel \textbf{\textcolor{blue}{\Comment{Pessimistic case}}}
			\begin{enumerate}
			    \item Compute $(y_{r,j}, \alpha_{r,j}) \leftarrow \texttt{VDF.Eval}(\textsf{pp}_j, x_{r,j}, T)$.
                \item Compute $\pi_{r,j} \leftarrow \texttt{VDF.Prove}(\textsf{pp}_j, x_{r,j}, \mu_r, y_{r,j}, \alpha_{r,j}, T)$.
                \item Compute $y_r = \prod_{P_i \in \mathcal{P}} y_{r, i} \cdot \prod_{P_j \in \tilde{\mathcal{P}}} y_{r, j}$
                \item Optionally, a proof $\pi_{y_r}$ can be computed to enable verification of $y_r$.
			\end{enumerate}
			\dotfill 
			
			\item Output the $r$-th round's randomness beacon $R_r \leftarrow \mathcal{H}_{\textsf{inputTorand}}(y_r)$
			\item \textbf{Reward the participants computing the $r$-th round's randomness beacon value $\mathcal{P} \setminus \tilde{\mathcal{P}}$ and apply a Penalty to  participants $\tilde{\mathcal{P}}$}.
			\item Set $r \leftarrow r + 1$.
			\item Repeat from step 2 to step 11 -- to generate the next round's randomness.
		\end{enumerate}

\end{ProblemSpecBox}

\begin{theorem}\label{theorem:drb}
	Assuming that $\mathcal{H}_{\textsf{\emph{randToinput}}}$ and $\mathcal{H}_{\textsf{\emph{inputTorand}}}$ is the random oracle and \texttt{\emph{VDF}} is a trapdoor watermarkable VDF, then it holds that \emph{Algorithm~\ref{algo:ranbec}} is a DRB scheme.
\end{theorem}

The proof of Theorem~\ref{theorem:drb} is deferred to the full version.

\section{Future Work}

Existing collaborative DRB protocols experience challenges in inefficient communication complexity, which limits their scalability.

\textit{In the near future, we hope to construct a complexity-efficient collaborative DRB protocol.}


\IEEEtriggercmd{\enlargethispage{-5in}}
\bibliographystyle{IEEEtran}
\bibliography{poster-ref}


\end{document}